\documentclass{jfm}

\usepackage{graphicx}

\usepackage{natbib}
\usepackage{multirow}
\usepackage{booktabs}
\usepackage{amsmath}
\usepackage{amssymb}
\usepackage{color}
\usepackage{moreverb}
\usepackage[geometry]{ifsym}

\ifCUPmtlplainloaded \else
  \checkfont{eurm10}
  \iffontfound
    \IfFileExists{upmath.sty}
      {\typeout{^^JFound AMS Euler Roman fonts on the system,
                   using the 'upmath' package.^^J}%
       \usepackage{upmath}}
      {\typeout{^^JFound AMS Euler Roman fonts on the system, but you
                   dont seem to have the}%
       \typeout{'upmath' package installed. JFM.cls can take advantage
                 of these fonts,^^Jif you use 'upmath' package.^^J}%
      }
  \else
  \fi
\fi

\ifCUPmtlplainloaded \else
  \checkfont{msam10}
  \iffontfound
    \IfFileExists{amssymb.sty}
      {\typeout{^^JFound AMS Symbol fonts on the system, using the
                'amssymb' package.^^J}%
       \usepackage{amssymb}%
       \let\le=\leqslant  
       \let\ge=\geqslant  
      }{}
  \fi
\fi

\ifCUPmtlplainloaded \else
  \IfFileExists{amsbsy.sty}
    {\typeout{^^JFound the 'amsbsy' package on the system, using it.^^J}%
     \usepackage{amsbsy}}
    {\providecommand\boldsymbol[1]{\mbox{\boldmath $##1$}}}
\fi

\newsavebox{\astrutbox}
\sbox{\astrutbox}{\rule[-5pt]{0pt}{20pt}}

\title[Dependence of decaying homgeneous isotropic turbulence on
  initial conditions]{Dependence of decaying homogeneous isotropic
  turbulence on initial conditions}

\author[P. C. Valente and J.C. Vassilicos]%
{P.\ns C.\ns V\ls A\ls L\ls E\ls N\ls T\ls E\break
J.\ns C.\ns V\ls A\ls S\ls S\ls I\ls L\ls I\ls C\ls O\ls S
 }

\affiliation{Department of Aeronautics, Imperial College London,
London SW7 2AZ, United Kingdom}

\pubyear{2011}
\volume{??}
\pagerange{??}
\date{?? and in revised form ??}

\begin{document}

\maketitle

\begin{abstract}
We conduct a careful analysis of the data provided by \cite{K&D2011}
and show that their data do not support their conclusions. According
to their published data, their decaying approximately homogeneous
isotropic turbulent flows are, invariably, clearly different from
Saffman turbulence; and very clearly marked
differences exist between the far downstream turbulence behaviors
generated by their conventional grid and by their multiscale cross
grids.
\end{abstract}

\section{Introduction} \label{sec:introduction}

A few years ago, \citet*{lavoie2007effects} investigated
potential effects of initial conditions on the decay of approximately
homogeneous isotropic turbulence. Initial conditions refer to the way
the turbulence is generated. In the wind tunnel experiments of these
authors, the turbulence was passively generated by square-mesh biplane
grids placed at the test section entry. A particular aspect of the
potential dependence on initial conditions is whether the power-law
decay of the far-downstream turbulence depends on
them. Quantitatively, the question is whether the decay exponent $n$
in
\begin{equation}
u^2 \sim (x-x_{0})^{-n}
\label{powerlaw}
\end{equation}
(where $u^{2}$ stands for two thirds of the turbulent kinetic energy
and $x$ is the streamwise distance along the tunnel, $x_0$ being a
virtual origin) differs for different initial conditions as claimed by
\cite{George1992}.

\cite{lavoie2007effects} tried four different conventional passive grids
(with square or with round bars with/without a small helical wire) and
two different test sections (one with and one without a secondary
contraction to improve isotropy). They did not find any significant
effect of initial conditions on the decay exponent $n$ other than that
of anisotropy which does, itself, depend on initial conditions and
persists far downstream.

\cite{K&D2011} carried out a similar wind tunnel study but with two
multiscale grids and one conventional grid. Their grids were all
monoplanar and their two multiscale grids were chosen from one of the
three families of multiscale grids introduced by \cite{H&V2007},
specifically the family of fractal cross grids. These grids are very
different from the low-blockage space-filling fractal square grids
which have been the multiscale grids of choice in the vast majority of
subsequent works on multiscale/fractal-generated turbulence
\citep*{S&V2007,nagata2008direct, nagata2008dns, SPSV2010, M&V2010,
  suzuki2010, L&V2011, V&V2011}.  The reason why multiscale/fractal
cross grids have mostly been neglected (except in studies where they
were used to enhance the Reynolds number, see \citet*{3DPTV,Lindstedt})
is that \cite{H&V2007} did not make any strong or unexpected claim
about the dependence of $u^{2}$ on $x-x_0$ in decaying turbulence
generated by them. Their conclusion on these grids was just a double
negative: ``the turbulence decay observed is not in disagreement with
power-law fits and the principle of large eddies''.

Sketches of the multiscale cross grids used by \cite{K&D2011} can be
seen in their figure 1 and are described in their section 2 where they
are labeled \emph{msg1} and \emph{msg2}. We do not need to repeat the
description here except to say that each multiscale cross grid has
three different mesh sizes, the smallest one being $M_3 = 15mm$ for
\emph{msg1} and $M_3 = 21mm$ for \emph{msg2}. \cite{K&D2011} were
careful to design their two multiscale cross grids and one
conventional grid in such a way that the longitudinal integral
length-scale of the turbulence at a 2m distance from the grid location
is the same $\ell_{0} \approx 23.65mm \pm 0.25mm$ for all three
grids. The ratio between $\ell_0$ and the distance between the tunnel
walls is smaller than $1/75$.

A description of the wind tunnel used by \cite{K&D2011} can be found
in \cite{K&D2010, K&D2011}. It is much larger and longer than the
tunnel used by \cite{H&V2007} and the grids were placed in the tunnel
contraction, specifically 1.2m upstream from the start of their test
section. As a result, the multiscale grid-generated turbulence of
\cite{K&D2011} is more isotropic and much further downstream of the
grid than in \cite{H&V2007}.  Their turbulence measurements were taken
using single and two component hot-wire anemometry from $x\approx 60
l_0$ till $x=400 \ell_0$ which means $93 M_{3}\le x \le 629 M_{3}$ for
\emph{msg1} and $67 M_{3}\le x \le 446 M_{3}$ for \emph{msg2}. This is
clearly much further downstream than \cite{H&V2007} who could not take
measurements beyond a distance equal to 80 times the smallest mesh
size of their own multiscale cross grids. In the case of
\cite{K&D2011} conventional grid (refered to as \emph{cg}), $60 \ell_0
\le x \le 400 \ell_0$ corresponds to $40 M \le x \le 240 M$ where $M$
is the mesh size of the grid.

The first main conclusion of \cite{K&D2011} was that their ``results
are at odds'' and that their ``findings contradict'' those of
\cite{H&V2007}. This claim is factually incorrect not only because the
multiscale cross grids in these two studies have significant
differences (different blockages, but also differences in some other
grid-defining parameters, see figure 1 in \cite{K&D2011} and figure 3
in \cite{H&V2007} and related parameters), but also because the
regions of the flow were different, in fact not even overlapping (if
just barely in one case), in terms of multiples of the smallest mesh
size of the multiscale cross grids. The fact that \cite{H&V2007}
measured higher turbulence levels and local Reynolds numbers
$Re_{\lambda} = u\lambda/\nu$ (where $\lambda$ is the Taylor
microscale and $\nu$ the kinematic viscosity) than \cite{K&D2011} is
quite simply consistent with the fact that \cite{H&V2007} measured
much closer to the grid. \cite{H&V2007} also extracted decay exponents
$n$ from their data and showed how they vary continuously as the
choice of virtual origin $x_0$ varies (see their figure 10). Choosing
$x_0$ close to $0$, they found exponents $n$ close to $1.2$ for their
multiscale cross grids and close to $1.4$ for their conventional grid
with large mesh size. However, they also proposed a $\lambda$-based
method for choosing $x_0$ which returned $n\approx 1.75$ for their
multiscale cross grids, $n\approx 2.3$ for their conventional grid
with high mesh size and $n\approx 1.39$ for their conventional grid
with usual mesh size. These particular $\lambda$-based exponents are
indeed higher than those claimed by \cite{K&D2011} which lie between
$1.12$ and $1.25$ but they were obtained in completely different
regions of the flow and, most importantly, with different fitting
methods which yielded very different values of $x_0$.
In fact \cite{H&V2007} did not include these exponents in their
conclusions because ``more extensive checks of large-scale and
small-scale isotropy as well as homogeneity will be required to fully
conclude on the nature of the turbulence decay behind fractal cross
grids, in particular in order to assess the viability of our
$\lambda$-based method for estimating $x_0$ in the cross
grid-generated flows''.

The second main conclusion of \cite{K&D2011} is that, in the
far-region where they measure, their multiscale cross grids and their
conventional grid produce ``virtually identical'' turbulence
behavior. Furthermore, quoting from their conclusion, ``Saffman's
decay law is reasonably robust, since the energy decay exponents for
all three grids are close to Saffman's classical prediction of
$n=6/5$''. In the next two sections we use the data published by
\cite{K&D2011} and show that an attentive analysis of their data based
on \cite{K&D2010} leads to very different conclusions.

\section{Decaying homogeneous isotropic turbulence with three different
initial conditions} \label{sec:decay}

\cite{K&D2011} established that their turbulent flows were reasonably
homogeneous at $x$ beyond $2m$ in terms of longitudinal profiles of
variances, skewnesses and flatnesses of the streamwise fluctuating
velocity component. Their centreline mean streamwise flow $U$ remains
constant to within less than $\pm 0.1 \%$ for all three grids from
$x=2m$ till about $x=8m$, though it deviates a very little bit for
\emph{msg2} beyond $x=6.5m$. As a result, they chose to design their
three grids in such a way that they all generate turbulence with
nearly same longitudinal integral length-scale $\ell_0$ at $x=2m$. The
positions $x=80\ell_0$ fall around $1.9m$ for all the
grids. (\cite{K&D2011} in fact recorded, and in a few instances used
for their analysis, a few measurements at closer distances to the
grid, i.e. $x$ as small as about $41 \ell_0$.) The longitudinal
length-scale $\ell$ grows as the turbulence moves downstream, but the
ratio between $\ell$ and the distance between the tunnel walls remains
very small, for example less than about $1/40$ at about $8m$ from the
grid location.

They also calculated ratios $<u_x^{2}>/<u_y^{2}>$,
$<u_x^{2}>/<u_z^{2}>$ and $u^{2}/<u_x^{2}>$ and found small levels of
anisotropy ``comparable, if not better, than in most other
experiments''. In particular, $u^{2}/<u_x^{2}>$ hovers between $0.95$
and $1.02$ throughout the regions where they recorded their
measurements. Hence any anisotropy-related dependence on initial
conditions as in \cite{lavoie2007effects} can, most probably, be ruled
out.

\begin{figure}
\centering
\includegraphics[trim=1mm 0mm 11mm 0,
clip=true,width=2.6in]{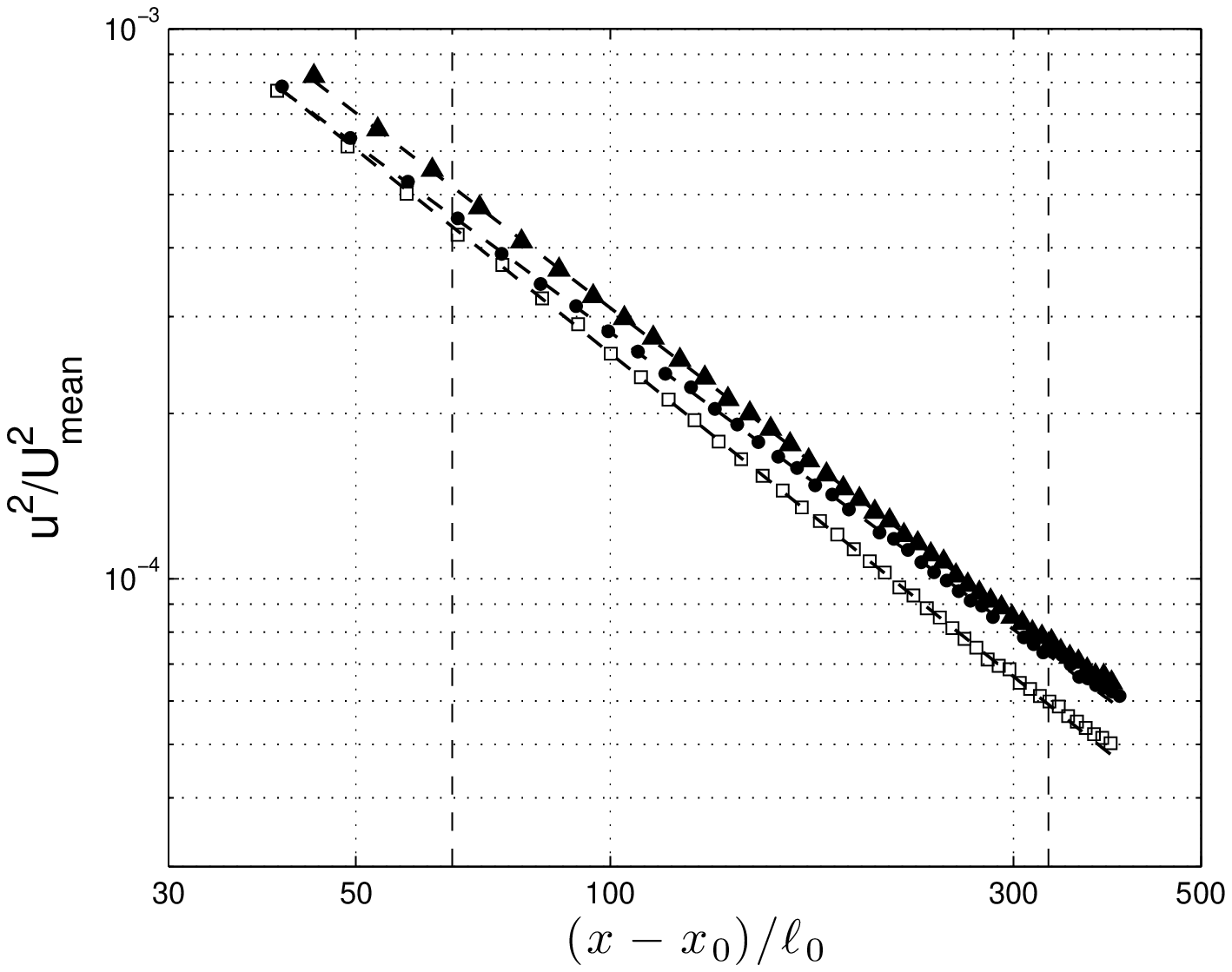}
\includegraphics[trim=1mm 0mm 7mm 0,
clip=true,width=2.6in]{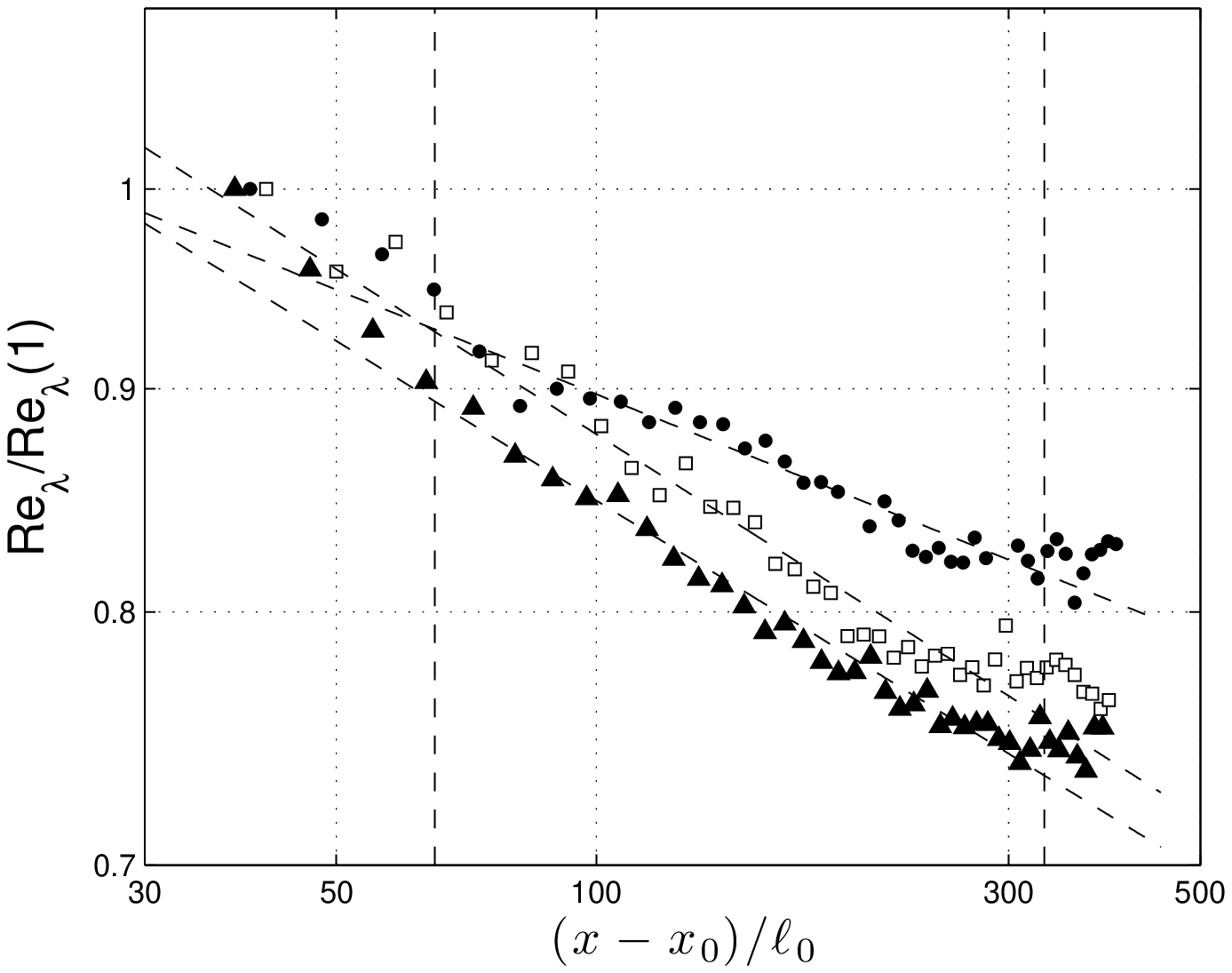}
\caption{Data and best-fit power laws for: (a)
  $<u_{x}^{2}>/U_{mean}^2$ versus $(x-x_0)/\ell_0$ (b)
  $Re_{\lambda}/Re_{\lambda 1}$ versus
  $(x-x_0)/\ell_0$. (\FilledSmallCircle) \emph{cg}, (\FilledTriangleUp
  ) \emph{msg1}, (\SmallSquare) \emph{msg2}. The vertical dashed lines
  mark the start and end of the admissible data range used in the
  least-squares fits.}
\label{Fig:DecayFit}
\end{figure}

In figure \ref{Fig:DecayFit}a we plot $<u_{x}^{2}>/U^{2}$ versus
$(x-x_{0})/\ell_0$ for all three grids as well as fits of the data by
$<u_{x}^2>/U^{2} \sim ({x-x_{0}\over \ell_{0}})^{-n}$. The decay
exponents $n$ and virtual origins $x_0$ in these fits are estimated
simultaneously by direct application of a non-linear least-squares
regression algorithm (`NLINFIT' routine in MATLAB$^{TM}$). This
fitting method is closely related to the one used by
\cite{lavoie2007effects} and we apply it to nearly the same range
where \cite{K&D2011} applied their own fitting methods. Specifically,
we apply our fit to the range $80 \ell_{0} < x < 330 \ell_{0}$ which is a
range of $x$ from about $1.9m$ to $8m$. This means that, for each
grid, we exclude data points obtained by \cite{K&D2011} at values of
$x$ smaller than $80\ell_{0}$ where according to these authors the
turbulence is not sufficiently homogeneous, and we also exclude,
exactly like \cite{K&D2011} do, the data points furthest downstream
where noise starts to be significant. (Including data points from
$x\approx 1.5m$ (i.e. $60\ell_0$) as in \cite{K&D2011} makes little
  difference as the values of $n$ remain the same to within $\pm
  0.01$.)

We give the values of $n$ and $x_0$ thus obtained in table
\ref{Table:ExponentAllRange} (method I). These values agree fairly
well with the various values of $n$ and $x_0$ obtained by
\cite{K&D2011} by their three different fitting methods for all three
grids except for their value of $x_0$ for \emph{msg1} and their value
of $n$ for \emph{msg1} when they use one of their three fitting
methods, the regression method (see their table 1). The values of $n$
which they obtain for \emph{msg1} with their other two fitting methods
are close to our value of $n$ for \emph{msg1}.

At this point it may be helpful to recall some basic theoretical
considerations. Homogeneous turbulence in the wind tunnel decays
according to $U {d\over dx} {3\over 2} u^{2} = -\epsilon$ where
$\epsilon$ is the turbulent kinetic energy dissipation per unit
mass. To obtain (\ref{powerlaw}) and the numerical value of $n$, one
needs some more information about $u^{2}$ and $\epsilon$. This
information usually consists of the following three ingredients when
the homogeneous turbulence can also be considered fairly isotropic
\cite[see][]{batchelor1948decay,Batchelor:book, Rotta:book}:
(i) a finite invariant of the von K\'arman-Howarth equation, (ii) the
assumption that the decay of large eddies is self-similar and (iii)
the empirical assumption that
\begin{equation}
A \equiv \epsilon \ell/u^{3}
\label{A}
\end{equation}
remains constant during decay ($ \ell = \ell(x)$ is the longitudinal
integral
length-scale). This constancy can be thought of as resulting from the
assumed independence of $A$ on turbulence intensity and
$Re_{\lambda}$.

\cite{V2011} proved that there are four different cases of finite
invariants of the von K\'arman-Howarth equation depending on
conditions at infinity. A case where no known finite invariant exists;
a case where the Loitsyansky invariant is the only known finite
invariant and where self-similar decay of large eddies implies
$u^{2}\ell^{5}=const$ during decay; a case where only one known finite
invariant exists and where self-similar decay of large eddies implies
$u^{2}\ell^{m+1}=const$ with $2\le m<4$ ($2\le m$ ensures that the
spectral tensor does not diverge at zero wavenumber as stated in
\cite{Rotta:book}, in the Appendix of \cite{K&D2011} and in
\cite{V2011}); and a case where two finite invariants exist and where,
as a consequence, self-similar decay of large-eddies is impossible.

Using the constancies of $A$ and $u^{2}\ell^{m+1}$, the second and
third of these four cases imply
\begin{equation}
n=2(m+1)/(m+3)
\label{nm}
\end{equation}
where $2\le m\le 4$ and therefore $6/5\le n\le 10/7$. There is no
known way to rule out the first and fourth cases and therefore no
known theoretical reason for measured values of $n$ to necessarily lie
inside the range $6/5\le n\le 10/7$.

Two out of the three present grids have returned values of $n$ which
are below $6/5=1.2$ (see table \ref{Table:ExponentAllRange} under
method I). However, this does not imply that the present turbulence
measurements do not fall under the second or third cases identified by
\cite{V2011}. Indeed, as \cite{K&D2010, K&D2011} have observed, $A$
varies slowly with $x$ and is therefore not strictly constant. If this
is so, then (\ref{nm}) needs to change.

\begin{figure}
\centering
\includegraphics[trim=2mm 0mm 7mm 0,
clip=true,width=2.5in]{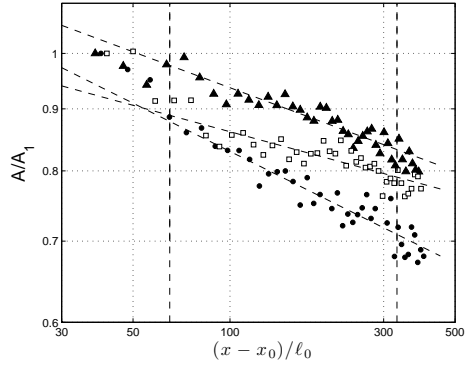}
\caption{Data and best-fit power laws for $A/A_{1}$ versus
  $(x-x_0)/\ell_0$. (\FilledSmallCircle) \emph{cg}, (\FilledTriangleUp
  ) \emph{msg1}, (\SmallSquare) \emph{msg2}. The vertical dashed lines
  mark the start and end of the admissible data range used in the
  least-squares fits.}
\label{Fig:CepsRelambdaNorm}
\end{figure}

In figure \ref{Fig:CepsRelambdaNorm} we plot the values of $A$
obtained by \cite{K&D2011} for their three grids as functions of
$(x-x_{0})/\ell_{0}$ where $x_0$ is taken from table
\ref{Table:ExponentAllRange} (method I). (\cite{K&D2011} assumed
small-scale isotropy and calculated $A$ from measurements of
$<({\partial u_{x}\over \partial x})^{2}>$ using $\epsilon = 15\nu
<({\partial u_{x}\over \partial x})^{2}>$ and integrations of measured
longitudinal correlation functions
to educe $\ell$.) To bring out more clearly the differences between
grids we in fact plot $A/A_1$ where $A_1$ is the value of $A$ obtained
at the smallest distance $x$ from each grid. We then follow
\cite{K&D2010} and fit the power law $A \sim ({x-x_{0}\over
  \ell_{0}})^{-p}$ in the range $60 \ell_{0} < x < 330 \ell_{0}$
of this data. These fits are shown in figure
\ref{Fig:CepsRelambdaNorm} and the values of $p$ are reported in
table \ref{Table:ExponentAllRange}.

If $A=const$ is replaced by $A \sim (x-x_{0})^{-p}$ then the
implication of $u^{2}\ell^{m+1}=const$ changes from (\ref{nm}) to
\begin{equation}
n=(1-p) 2(m+1)/(m+3)
\label{nmcorr}
\end{equation}
where $2\le m\le 4$. With our estimates of $n$ and $p$ we can now
use (\ref{nmcorr}) to derive values of $m$ for each grid. They are
given in table \ref{Table:ExponentAllRange} (under method I) and,
having now taken into account the slight variations of $A$, they are
all between $2$ and $4$. Similarly, the values of $n_{corr} \equiv
n/(1-p)$ lie all between $6/5$ and $10/7$ (see table
\ref{Table:ExponentAllRange} under method I).

\begin{figure}
\centering
\includegraphics[trim=16mm 0mm 7mm 0, 
clip=true,width=2.6in]{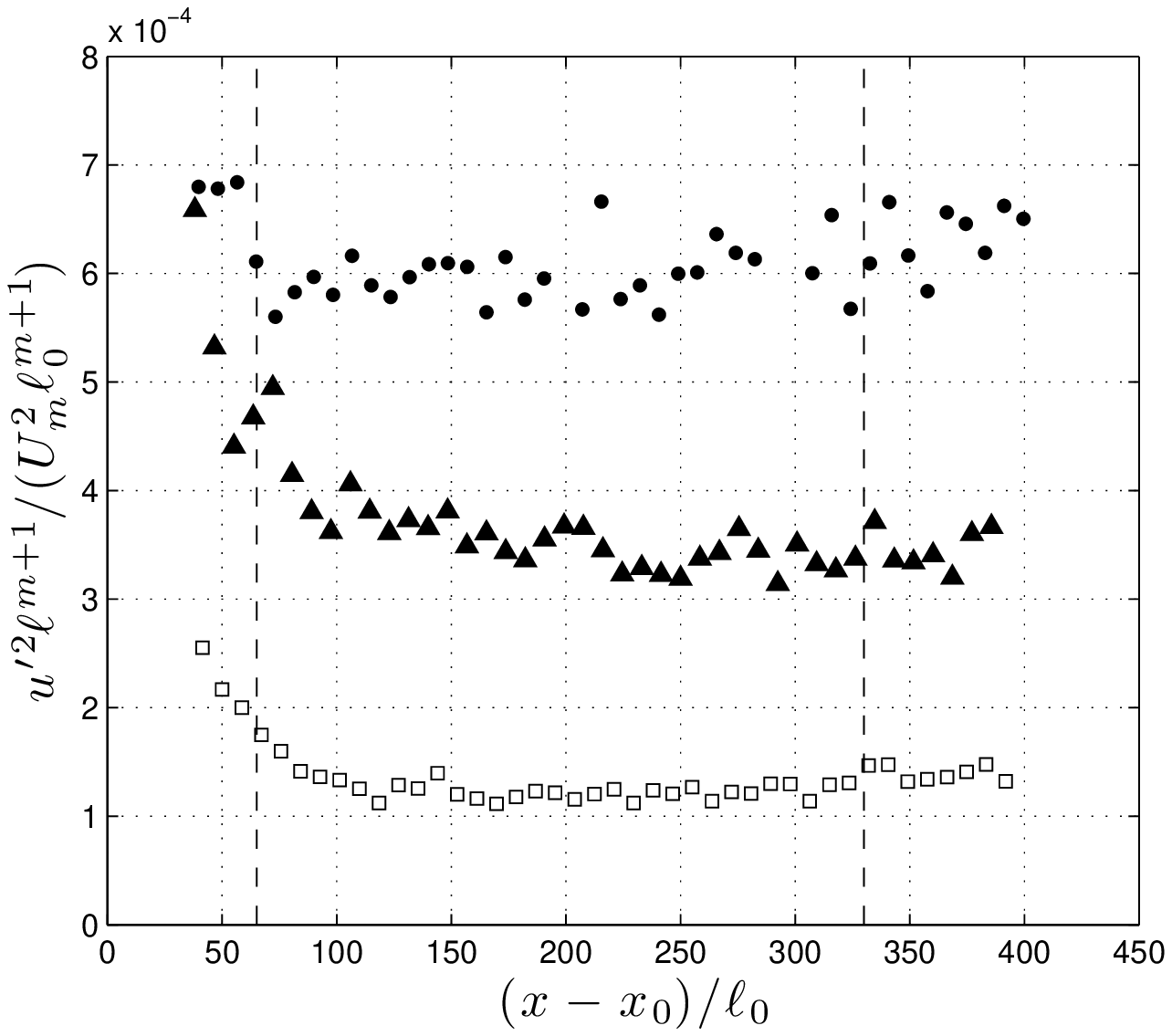}
\includegraphics[trim=16mm 0mm 7mm 0,
clip=true,width=2.6in]{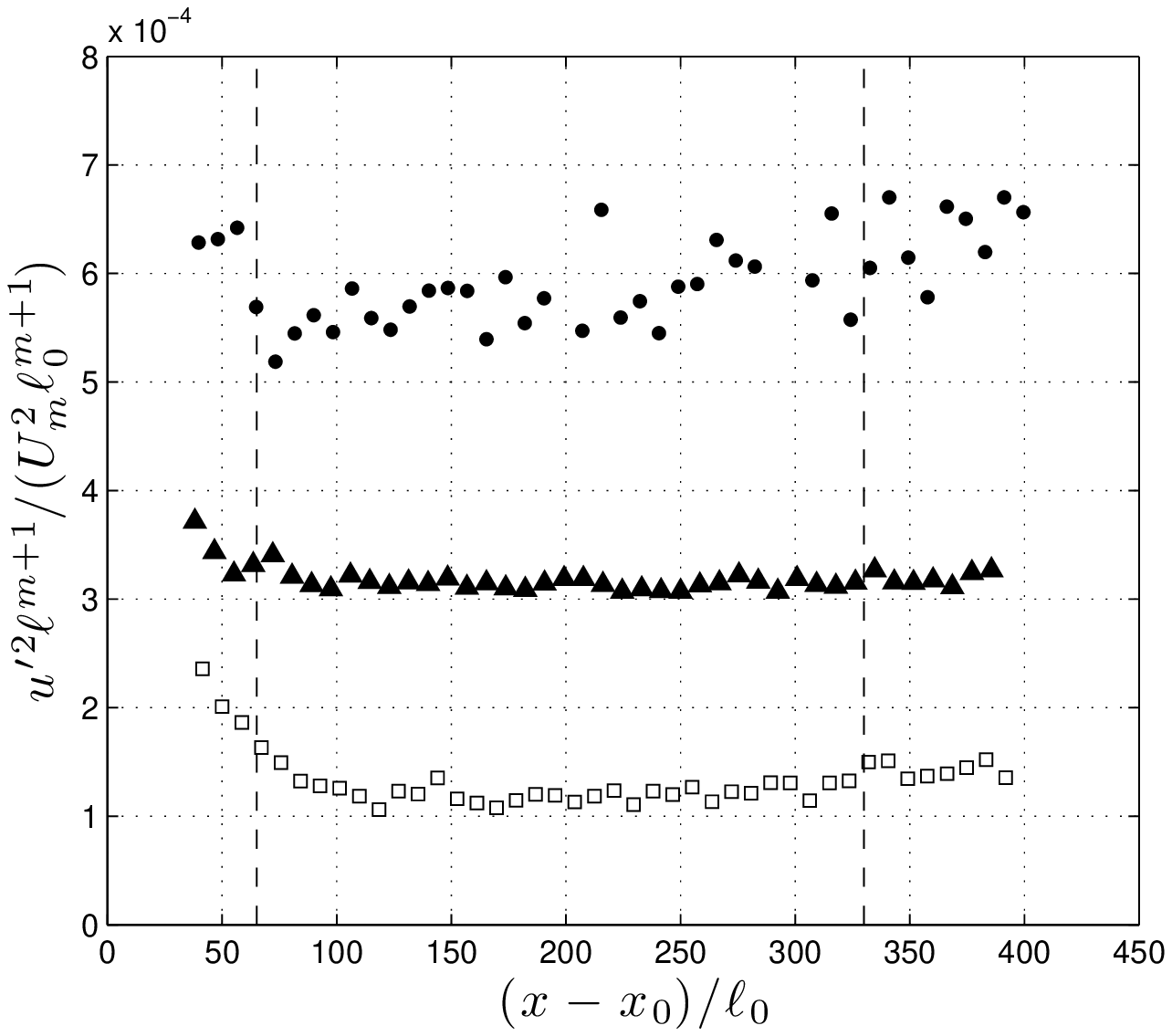}
\includegraphics[trim=10mm 0mm 12mm 0,
clip=true,width=2.6in]{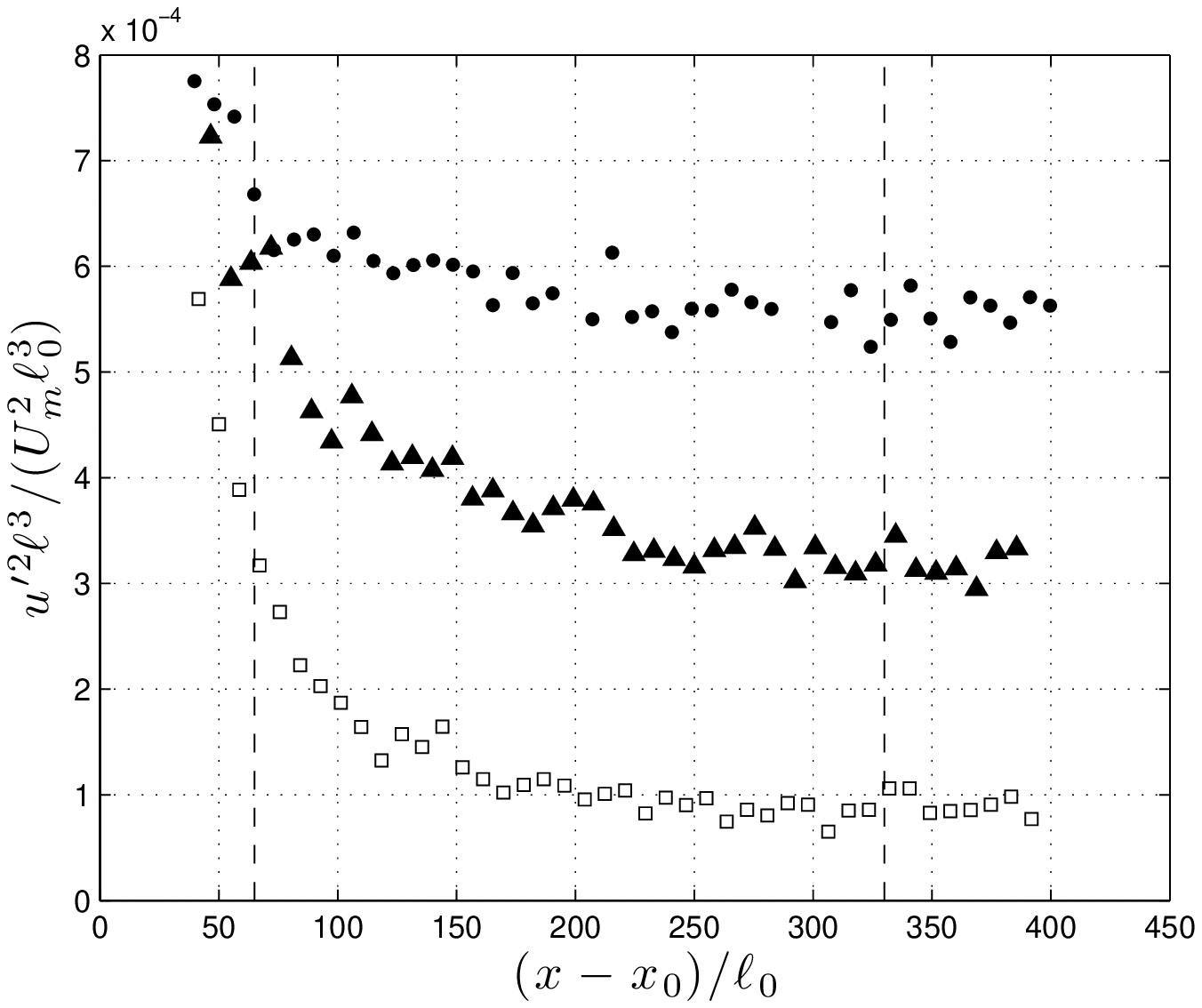}
\caption{Checks of invariant forms via plots of $u^2\ell^{m+1}$: (a)
  $m$ from method I, (b) $m$ from method II, (c) $m=2$ corresponding
  to Saffman turbulence. (\FilledSmallCircle) \emph{cg},
  (\FilledTriangleUp ) \emph{msg1}, (\SmallSquare) \emph{msg2}. For
  improved readability the \emph{cg}/\emph{msg1}/\emph{msg2} data were
  vertically offset by (a) $[2.0,\, 1.0,\, 0.0] \times10^{-4}$, (b)
  $[1.6,\, 2.5,\, 0.0] \times10^{-4}$, (c) $[2.5,\, 0.5,\,
    -1.5]\times10^{-4}$.The left and right vertical dashed lines mark
  the start and end of the data range used to obtain the decay
  exponents $n$. This is the range not significantly affected by
  inhomogeneity (to the left) and noise (to the right).}
\label{Fig:Invariants}
\end{figure}

These values of $m$ raise the possibility that the three decaying
nearly homogeneous and nearly isotropic turbulent flows of
\cite{K&D2011} may be three different instances of the third case
identified by \cite{V2011} where only one known finite invariant
exists and where self-similar decay of large eddies implies
$u^{2}\ell^{m+1}=const$ with $2\le m < 4$. The Saffman invariant
corresponds to $m=2$ but none of the grids used by \cite{K&D2011}
returns such a value of $m$. In figure \ref{Fig:Invariants}a we plot
$<u_{x}^{2}>\ell^{m+1}/(U^{2}\ell_{0}^{m+1})$ versus
$(x-x_{0})/\ell_{0}$ with the values of $m$ given under method I in
table \ref{Table:ExponentAllRange} for each one of the three different
turbulent flows. This figure should be compared with figure
\ref{Fig:Invariants}c which is a reproduction of figure 10 in
\cite{K&D2011} where they plotted
$<u_{x}^{2}>\ell^{3}/(U^{2}\ell_{0}^{3})$ versus $(x-x_{0})/\ell_{0}$,
except that we have offset the data vertically so as to see more
clearly the differences in behavior between each grid. Assuming the
turbulence is sufficiently homogeneous and isotropic and equally so
for all three flows \cite[as claimed by][]{K&D2011}, it is clear that
the Saffman prediction is not satisfied in these flows. Instead,
\begin{equation}
u^{2}\ell^{m+1}=const
\label{inv}
\end{equation}
with $m>2.5$ for all grids in the range $100 \ell_{0}\le x-x_{0}\le
400\ell_{0}$. Furthermore, different grids give rise to different
values of $m$ reaching up to $m\approx 3$ with method I (see table
\ref{Table:ExponentAllRange}).

In fact there is another way to extract values for $n$ and $m$ from
the data (method II), and this way gives even better defined
invariants and even greater differences between the far downstream
turbulence decays originating from the conventional grid and the
multiscale cross grids. Method II is based on figure 1b. This figure
is a log-log plot of $Re_{\lambda}/Re_{\lambda 1}$ versus
$(x-x_{0})/l_{0}$ where $Re_{\lambda 1}$ is the value of
$Re_{\lambda}$ at the smallest distance from each grid on this plot
and $x_0$ is the virtual origin obtained from our nonlinear fit of
figure 1a. The first inescapable observation is that the streamwise
distributions of $Re_{\lambda}$ are clearly different for the
conventional grid and for the multiscale grids.

The power law form (\ref{powerlaw}) implies $\lambda^{2} \sim
(x-x_{0})$ in decaying homogeneous isotropic turbulence
(\cite{Batchelor:book}). It follows that $Re_{\lambda} \sim
(x-x_{0})^{(1-n)/2}$, so that a best fit of the data in figure 1b
gives values of $n$. We apply this power law fit to the very same
range $80l_{0} < x < 330l_{0}$ used in method I for our fit of the
turbulence intensity data in figure 1a. The values of $n$ thus
obtained, the resulting values of $m$ using (\ref{nmcorr}) and the
resulting $n_{corr} \equiv n/(1+p)$ are given in table
\ref{Table:ExponentAllRange} under method II. In figure
\ref{Fig:Invariants}b we use these new values of $m$ to plot
$<u_{x}^{2}>\ell^{m+1}/(U^{2}\ell_{0}^{m+1})$ versus
$(x-x_{0})/\ell_{0}$ and find that they yield even better defined
invariants (\ref{inv}) than method I (compare with figure
\ref{Fig:Invariants}a). The difference between values of $m$ for
conventional grids and values of $m$ for multiscale grids is
unmistakable and even greater with method II than with method I.

We must conclude that the decay of approximately homogeneous
turbulence far from its initial conditions remains dependent on these
initial conditions. The decay exponent $n$ and the conserved finite
invariant $u^{2}\ell^{m+1}$ both clearly change when the
turbulence-generating grid is changed. These initial conditions may
have to do with the geometry of the grids or/and with the inlet
Reynolds numbers as the mean speed in the tunnel was $13.5m/s$ when
the conventional grid was tested, $14.0m/s$ when \emph{msg1} was
tested and $15.5m/s$ when \emph{msg2} was tested.

In trying to identify the flow-relevant geometrical variations from
grid to grid, we note that the mustiscale grids of \cite{K&D2011} have
three different bar widths $t_1 = 8mm$, $t_2 =4mm$ and $t_3 = 2mm$ as
well as three different mesh sizes $M_1 =2M_2$ and $M_2 \approx 2M_3$
with $M_1 = 64mm$ in the case of \emph{msg1} and $M_1 = 88mm$ in the
case of \emph{msg2} (see their figure 1). Hence the ratio of mesh size
to bar width is 11 for \emph{msg2} and 8 for \emph{msg1}. It is
therefore at least double than $M/t = 4$ for the conventional grid
\emph{cg}.

The mesh size determines the distance between the wakes of the bars
and the bar thickness determines the width of these wakes. Hence the
ratio of mesh size to bar thickness determines the distance from the
grid where the wakes meet and this distance increases when we move
from \emph{cg} to \emph{msg1} and \emph{msg2}.

The Reynolds numbers characterising these wakes (calculated as the
mean flow speed multiplied by the bar thickness and divided by the
kinematic viscosity of the air) take the values $3.6 \times 10^{3}$ in
the case of \emph{cg}; $3\times 10^{3}$, $1.5 \times 10^{3}$ and
$7.5\times
10^{2}$ in the case of \emph{msg1}; and $3.32\times 10^{3}$, $1.66 \times
10^{3}$ and $8.3\times 10^{2}$ in the case of \emph{msg2}. Unlike
conventional grids, multiscale grids impose more than one Reynolds
number on the flows they generate and a number of different distances
from the grid where wakes of different sizes meet. Of course the
largest wakes are affected by the wakes generated by the smaller ones.
But it is clear that the turbulence undergoes different generation
mechanisms extending over different streamwise distances with
different grids.
It is indeed remarkable that memory of these mechanisms remains in the
values of $n$ and $m$ as far downstream as where \cite{K&D2011} took
their measurements.

\begin{table}
\centering
  \caption{Estimation of quantities via least squares fit }
  \begin{tabular*}{0.8\textwidth}{@{\extracolsep{\fill}}c c ccccc cccc}
           &    & \multicolumn{5}{c}{Method I} &  \multicolumn{4}{c}{Method II} \\
           \cline{3-7}  \cline{8-11}\\
   Grid & $p$ & $n$ & $x_0$(m) & $m$ & $n_{corr}$ & $\alpha$ & $n$ & $m$ & $n_{corr}$ & $\alpha$ \\
    \midrule
    \emph{cg}      & 0.126 & 1.13 & 0.23 &  2.67 & 1.29 & 1.90    & 1.15 &  2.85 & 1.32 & 1.68\\ 
    \emph{msg1} & 0.101 & 1.18 & 0.28 &  2.79 & 1.31 & 1.14    & 1.24 &  3.38 & 1.37 & 0.86\\ 
    \emph{msg2} & 0.072 & 1.23 & 0.33 &  2.94 & 1.33 & 0.62    & 1.25 &  3.14 & 1.35 & 0.57\\
    \hline
  \end{tabular*}
\label{Table:ExponentAllRange}
\end{table}

\section{Different far-field low-$Re_{\lambda}$ turbulent flows}
\label{sec:decay}

\begin{figure}
\centering
\includegraphics[trim=0mm 0mm 10mm 0, clip=true,width=2.5in]{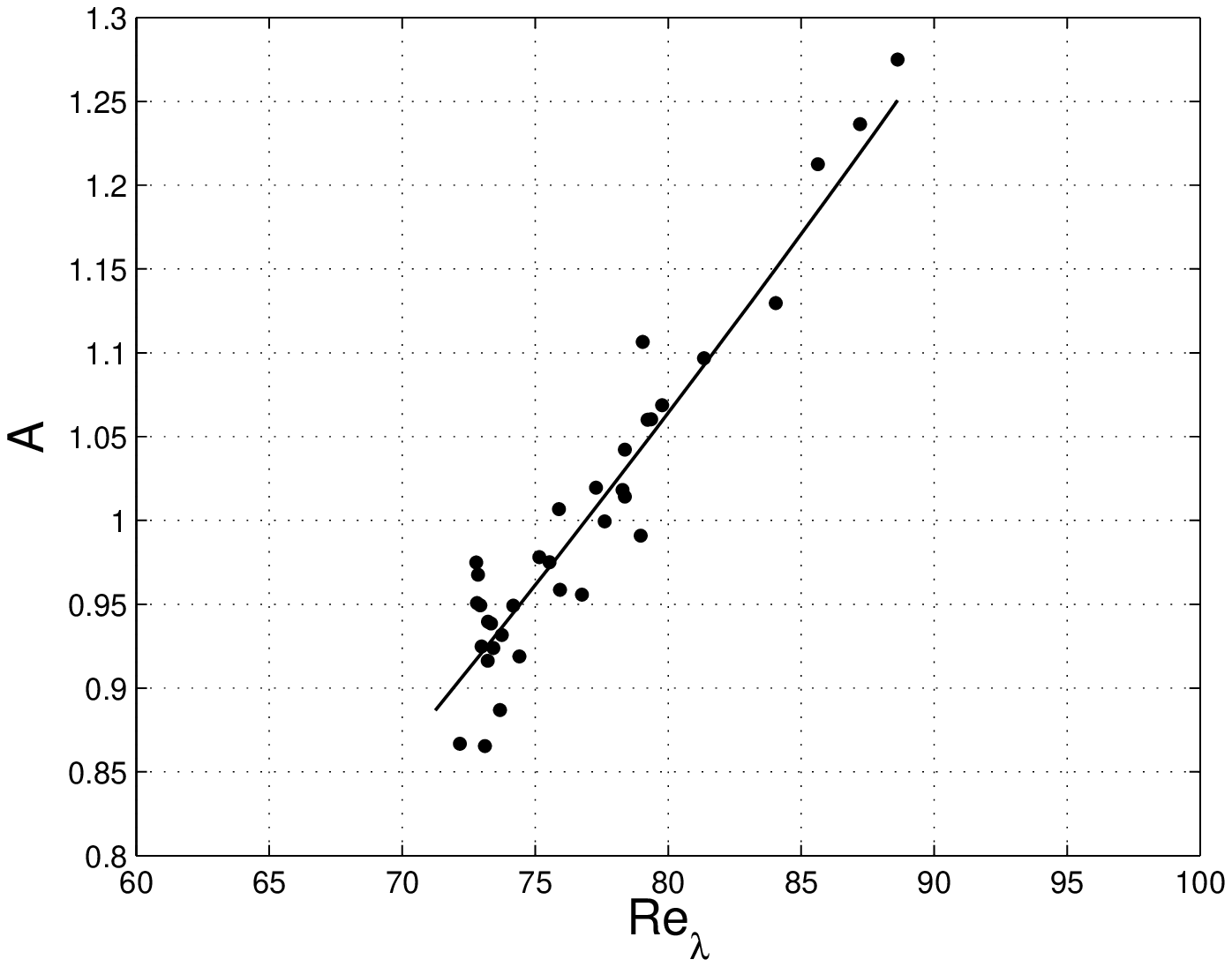}
\includegraphics[trim=0mm 0mm 10mm 0, clip=true,width=2.5in]{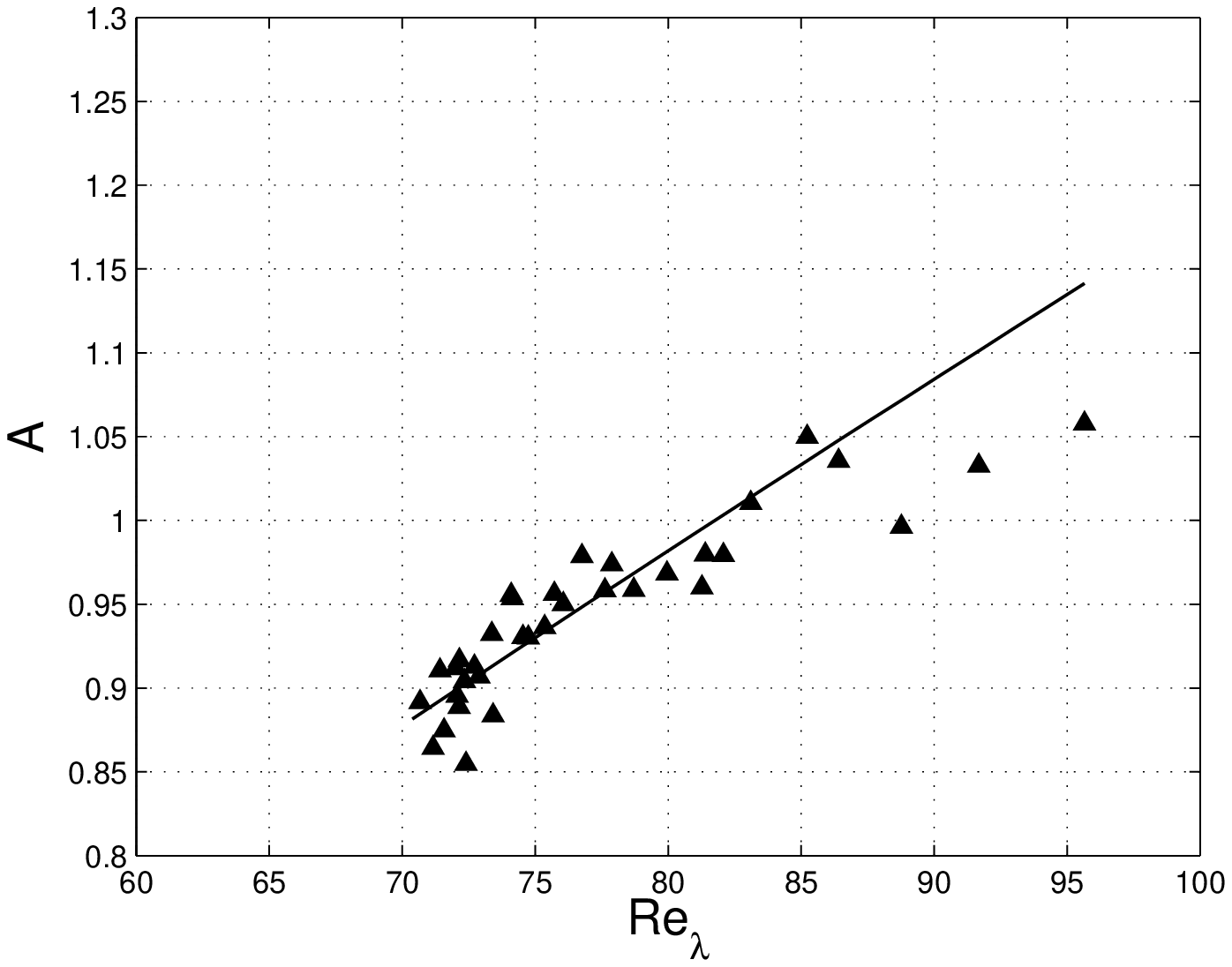}
\includegraphics[trim=0mm 0mm 10mm 0, clip=true,width=2.5in]{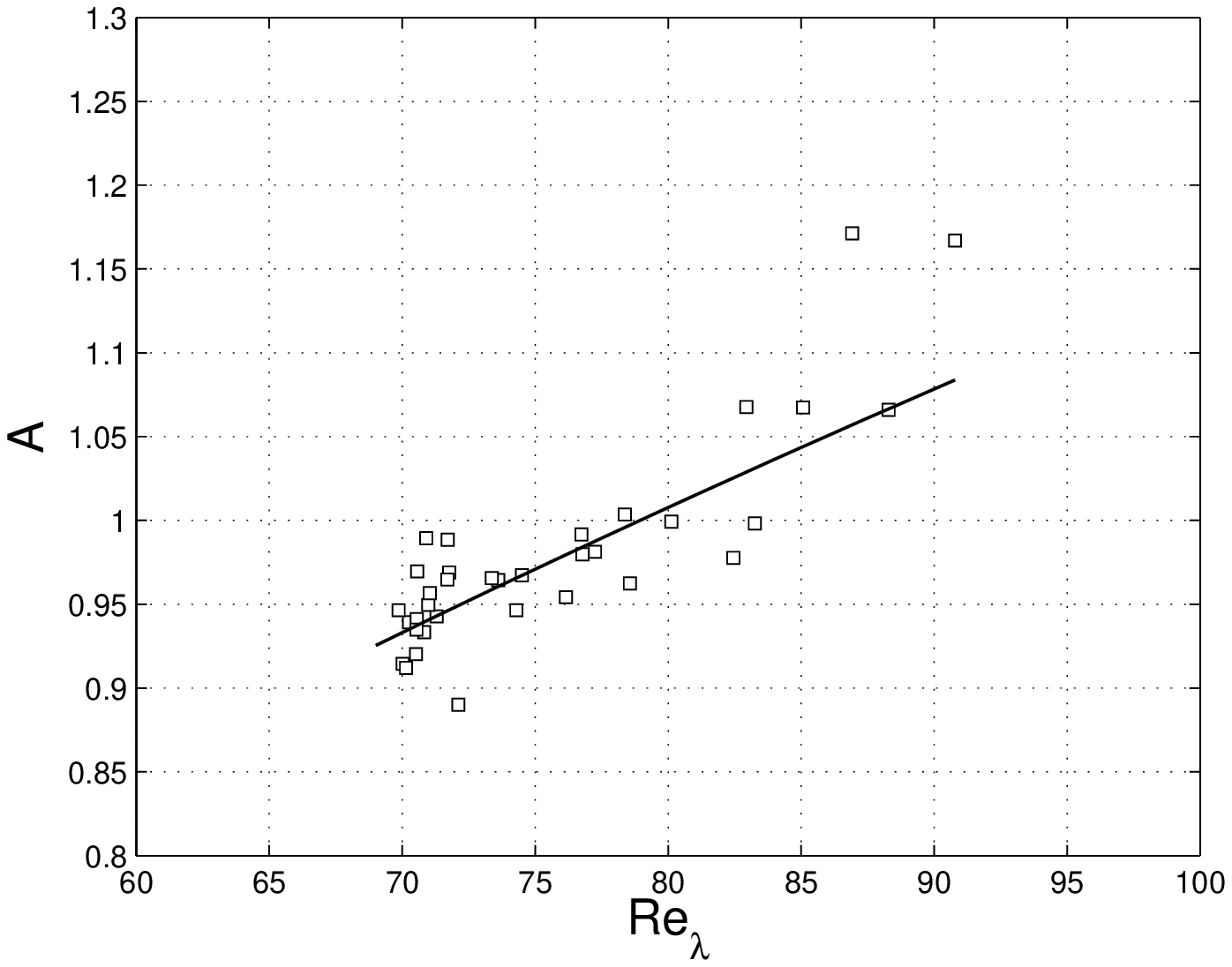}
\includegraphics[trim=0mm 0mm 10mm 0, clip=true,width=2.5in]{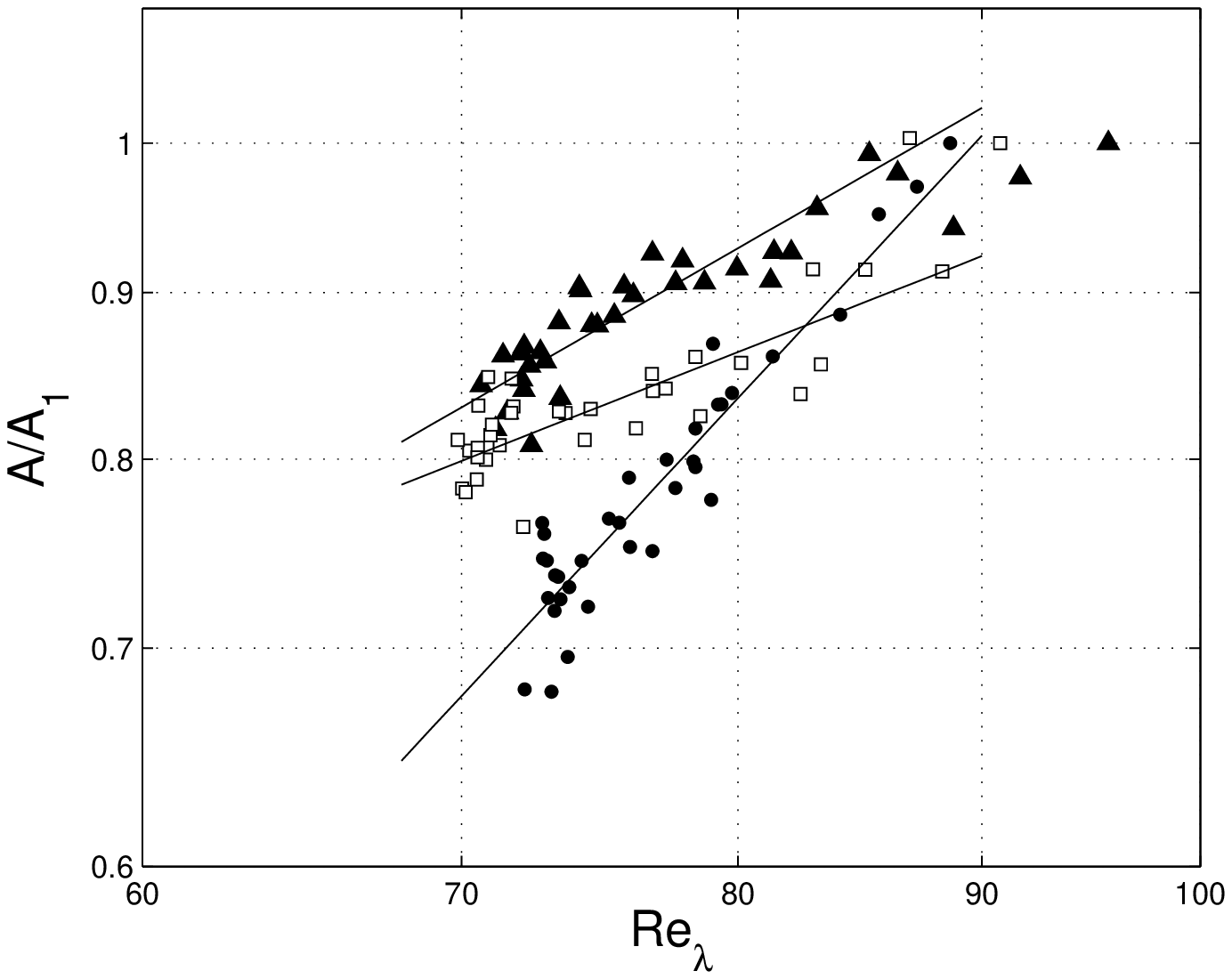}
\caption{$Re_{\lambda}$ dependence of the normalised energy
  dissipation rate A (note that $A$ here is $3A/2$ in \cite{K&D2011},
  for example see their figure 11): (a) \emph{cg}, (b) \emph{msg1},
  (c) \emph{msg2}, (d) logarithmic axes.  (\FilledSmallCircle)
  \emph{cg}, (\FilledTriangleUp) \emph{msg1}, (\SmallSquare)
  \emph{msg2}. The solid lines are plots of $A = const \times
  Re_{\lambda}^{\alpha}$ with $\alpha$ taken from table
  \ref{Table:ExponentAllRange}, method II.}
\label{Fig:CepsRelambda}
\end{figure}

We obtained figures \ref{Fig:Invariants}a,b by taking into account the
slow streamwise variation of $A$ as suggested by \cite{K&D2010}. This
streamwise variation can result from the well-known dependence that
the dimensionless dissipation rate $A$ has on $Re_{\lambda}$ when
$Re_{\lambda}$ is below at least 100
\cite*[e.g.][]{Burattini2005}. Indeed, the values of $Re_{\lambda}$
characterising the three far-field turbulent flows of \cite{K&D2011}
range between about $90$ near $x\approx 60 \ell_{0}$ and $70$ at
$x\approx 330 \ell_{0}$. Using $Re_{\lambda} \sim (x-x_{0})^{(1-n)/2}$
and $A\sim (x-x_{0})^{-p}$ we obtain
\begin{equation}
A\sim Re_{\lambda}^{\alpha}
\label{ARe}
\end{equation}
where
\begin{equation}
\alpha = 2p/(n-1).
\label{alpha}
\end{equation}
The values of $\alpha$ implied by this formula on the basis of the
exponents $p$ and $n$ obtained in the previous section are very
different for different grids, ranging from $\alpha\approx 1.7$ to
$\alpha \approx 0.6$ (using $n$ obtained from method II, see table
\ref{Table:ExponentAllRange}). Figure \ref{Fig:CepsRelambda} confirms
how dramatically different the dependencies of $A$ on $Re_{\lambda}$
are for the multiscale grids and for the conventional grid.
Hence, multiscale grids definitely do not ``produce almost identical
results to the equivalent classical grids'' as claimed by
\cite{K&D2011}.

Figure \ref{Fig:CepsRelambda} also shows that
(\ref{ARe})-(\ref{alpha}) give rise to more or less reasonable fits of
the data thus lending support to the idea that much of the streamwise
variation of $A$ comes from its dependence on
$Re_{\lambda}$. Increasing values of the dimensionless dissipation
rate $A$ with increasing $Re_{\lambda}$ have also been reported in
previous works with square bar grids at such relatively low Reynolds
numbers, see for example figure 1 in \cite{Burattini2005}, table 4 in
\citet{CC71} and table 3 in \citet{GC74}.

\section{Conclusion} \label{sec:conclusion}

According to the published data in \cite{K&D2011}, multiscale cross
grids and their equivalent (in terms of $\ell_0$) conventional grid
can produce very different far-field approximately homogeneous
isotropic turbulence with wide variations in the dimensionless
dissipation rate's dependence on $Re_{\lambda}$. This would seem to
confirm the observation already made by \cite{Burattini2005} on the
basis of different $Re_{\lambda}$ dependencies of $A$ for different
grids, namely that ``the geometry of the grid appears to have a
persistent influence in the streamwise direction up to $x/M =80$''. In
fact the data of \cite{K&D2011} extend this observation to much
further distances downstream and to a wider range of grids.

This data also leads to the conclusion that the decay of the three
approximately homogeneous isotropic turbulent flows of \cite{K&D2011}
is characterised by an invariant quantity $u^{2}\ell^{m+1}$ in the
region of the flow $x \ge 80 \ell_{0}$ which is the most clearly
homogeneous. The exponent $m$ is significantly different from
Saffman's $m=2$ and ranges between 2.7 and 3.4 for the grids used by
\cite{K&D2011}. Their multiscale grids return values of $m$ which are
markedly larger than the values of $m$ returned by their conventional
grid. The streamwise distributions of $Re_{\lambda}$ and $A$ are also
very clearly different.

Finally we repeat the remark in our introduction that the data of
\cite{K&D2011} do not contradict previous findings on multiscale grids
but in fact complement them.

\begin{acknowledgments}
We thank Professor Per-\r{A}ge Krogstad for kindly providing us with
the post-processed data published in \cite{K&D2011}.
\end{acknowledgments}

\bibliographystyle{jfm}
\bibliography{VV2011JFM2_arxiv}

\end{document}